\def\year{2018}\relax
\documentclass[letterpaper]{article} 
\usepackage{aaai18}  
\usepackage{times}  
\usepackage{helvet}  
\usepackage{courier}  
\usepackage{url}  
\usepackage{graphicx}  
\usepackage{multirow}
\usepackage{hyperref}
\usepackage{lipsum}
\usepackage{graphicx,dblfloatfix}%
\usepackage{amsmath}
\begin{document}
%

\title{CrowdTruth 2.0: Quality Metrics for Crowdsourcing with Disagreement}
\author{
 Anca Dumitrache\thanks{Equal contribution, authors listed alphabetically.}, Oana Inel\footnote[1]{}, \\
 {\bf\Large Lora Aroyo} \\
 Vrije Universiteit Amsterdam \\
 \texttt{\{anca.dmtrch,oana.inel,} \\
 \texttt{l.m.aroyo\}@gmail.com} \\
 \And
 Benjamin Timmermans \\
 CAS IBM Nederland \\
 \texttt{b.timmermans@nl.ibm.com} \\
 \And
 Chris Welty\\
 Google Research, New York \\
 \texttt{cawelty@gmail.com} \\ 
}

\maketitle
\begin{abstract}
Typically crowdsourcing-based approaches to gather annotated data use inter-annotator agreement as a measure of quality. However, in many domains, there is ambiguity in the data, as well as a multitude of perspectives of the information examples. In this paper, we present ongoing work into the CrowdTruth metrics, that capture and interpret inter-annotator disagreement in crowdsourcing. The CrowdTruth metrics model the inter-dependency between the three main components of a crowdsourcing system -- worker, input data, and annotation. The goal of the metrics is to capture the degree of ambiguity in each of these three components. The metrics are available online at \url{https://github.com/CrowdTruth/CrowdTruth-core}.
\end{abstract}

\section{Introduction}

The process of gathering ground truth data through human annotation is a major bottleneck in the use of information extraction methods. Crowdsourcing-based approaches are gaining popularity in the attempt to solve the issues related to volume of data and lack of annotators. Typically these practices use inter-annotator agreement as a measure of quality. However, this assumption often creates issues in practice. Previous experiments we performed~\cite{aroyo2013crowd} found that inter-annotator disagreement is usually never captured, either because the number of annotators is too small to capture the full diversity of opinion, or because the crowd data is aggregated with metrics that enforce consensus, such as majority vote.  These practices create artificial data that is neither general nor reflects the ambiguity inherent in the data.

To address these issues, we proposed the {\bf CrowdTruth}~\cite{aroyo2014AIMag} method for crowdsourcing ground truth by harnessing inter-annotator disagreement.  We present an alternative approach for crowdsourcing ground truth data that, instead of enforcing agreement between annotators, captures the ambiguity inherent in semantic annotation through the use of disagreement-aware metrics for aggregating crowdsourcing responses.  In this paper, we introduce the second version of {\bf CrowdTruth metrics} -- a set of metrics that capture and interpret inter-annotator disagreement in crowdsourcing annotation tasks. As opposed to the first version of the metrics, published in~\cite{inel2014crowdtruth}, the current version models the {\it inter-dependency between the three main components of a crowdsourcing system -- worker, input data, and annotation}. This update is based on the intuition that disagreement caused by low quality workers should not be interpreted as the data being ambiguous, but also that ambiguous input data should not be interpreted as due to the low quality of the workers.

This paper presents the definitions of the CrowdTruth metrics 2.0, together with the theoretical motivations of the updates based on the previous version 1.0. The code of the implementation of the metrics is available on the CrowdTruth Github.\footnote{\url{https://github.com/CrowdTruth/CrowdTruth-core}} The 2.0 version of the metrics has already been applied successfully to a number of use cases, e.g. semantic frame disambiguation~\cite{dumitrache2018capturing}, relation extraction from sentences~\cite{dumitrache2017false}, topic relevance~\cite{inel2018studying}. In the future, we plan to continue the validation of the metrics through evaluation over different annotation tasks, comparing CrowdTruth approach with other disagreement-aware crowd aggregation methods.

\section{CrowdTruth Methodology}

 \begin{figure}[!bth]
 	\centering
 		\includegraphics[width=0.7\linewidth]{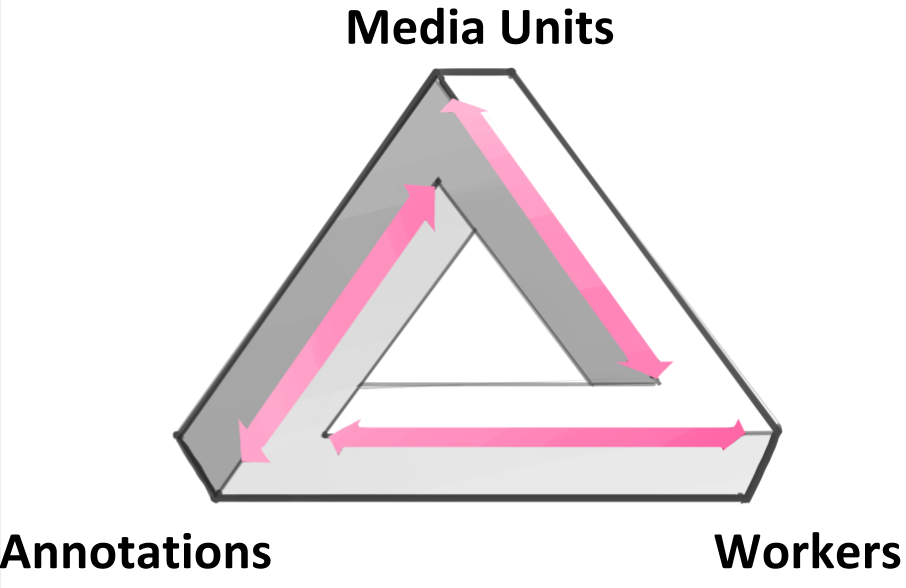}
 	\caption{Triangle of Disagreement}
 	\label{fig:triangle_of_reference}
 \end{figure}

\begin{figure*}[!tb]
 	\centering
 	\includegraphics[width=0.8\linewidth]{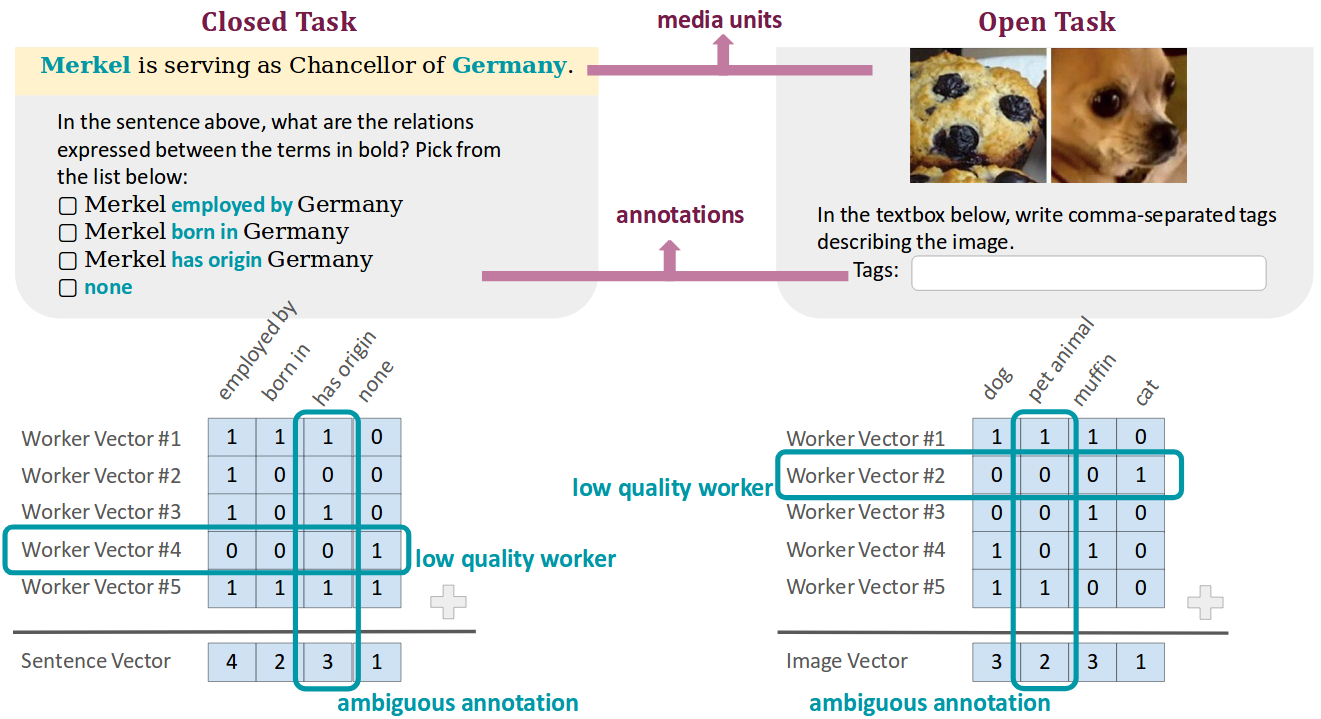}
 	\caption{Example closed and open tasks, together with the vector representations of the crowd answers.}
 	\label{fig:tasks}
 \end{figure*}

The CrowdTruth methodology consists of a set of quality metrics and best practices to aggregate inter-annotator agreement such that ambiguity in the task is preserved. The methodology uses the triangle of disagreement model (based on the triangle reference \cite{knowlton1966definition}) to represent the crowdsourcing system and its three main components -- input media units, workers, and annotations (Figure~\ref{fig:triangle_of_reference}). The triangle model expresses how ambiguity in any of the corners disseminates and influences the other components of the triangle. For example, an unclear sentence or an ambiguous annotation scheme would cause more disagreement between workers \cite{aroyo2014threesides}, and thus, both need to be accounted for when measuring the quality of the workers.

The CrowdTruth methodology calculates quality metrics for workers, media units and annotations. The novel contribution of version 2.0 is that {\it the way how ambiguity propagates between the three components of the crowdsourcing system has been made explicit in the quality formulas of the components}. So for example, the quality of a worker is weighted by the quality of the media units the worker has annotated, and the quality of the annotations in the task.

This section describes the two steps of the CrowdTruth methodology:
\begin{enumerate}
\item formalizing the output from crowd tasks into {\bf annotation vectors};
\item calculating quality scores over the annotation vectors using {\bf disagreement metrics}.
\end{enumerate}




\subsection{Building the Annotation Vectors}
\label{sec:vectors}

In order to measure the quality of the crowdsourced data, we need to formalize crowd annotations into a \textbf{vector space representation}. For \emph{closed tasks}, the annotation vector contains the given answer options in the task template, which the crowd can choose from. For example, the template of a \emph{closed task} can be composed of a multiple choice question, which appears as a list checkboxes or radio buttons, thus, having a finite list of options to choose from. Figure~\ref{fig:tasks} shows an example of a closed and an open task, indicating also what the media units and annotations are for both cases.

While for \emph{closed tasks} the number of elements in the annotation vector is known in advance, for \emph{open-ended tasks} the number of elements in the annotation vector can only be determined when all the judgments for a media unit have been gathered. An example of such a task can be highlighting words or word phrases in a sentence, or as an input text field where the workers can introduce keywords. In this case the answer space is composed of all the unique keywords from all the workers that solved that media unit. As a consequence, all the media units in a closed task have the same answers space, while for open-ended tasks the answer space is different across all the media units. Although the answer space for open-ended tasks is not known from the beginning, it still can be further processed in a finite answer space.

In the annotation vector, each answer option is a boolean value, showing whether the worker annotated that answer or not. This allows the annotations of each worker on a given media unit to be aggregated, resulting in a \textbf{media unit vector} that represents for each option how often it was annotated. Figure~\ref{fig:tasks} shows how the worker and media unit vectors are formed for both a closed and an open task.

\subsection{Disagreement Metrics}
\label{sec:metrics}

Using the vector representations, we calculate three core metrics that capture the {\bf media unit quality}, {\bf worker quality} and {\bf annotation quality}. These metrics are mutually dependent (e.g. the media unit quality is weighted by the annotation quality and worker quality), based on the idea from the triangle of disagreement that ambiguity in any of the corners disseminates and influences the other components of the triangle. The mutual dependence requires an iterative dynamic programming approach, calculating the metrics in a loop until convergence is reached. All the metrics have scores in the $[0,1]$ interval, with $0$ meaning low quality and $1$ meaning high quality. Before starting the iterative dynamic programming approach, the quality metrics are initialized with $1$. 

To define the CrowdTruth metrics, we introduce the following notation:
\begin{itemize}
\item $workers(u):$ all workers that annotate media unit $u$;
\item $units(i):$ all input media units annotated by worker $i$;
\item $WorkVec(i, u):$ annotations of worker $i$ on media unit $u$ as a binary vector;
\item $MediaUnitVec(s) = \sum_{i \in workers(s)} WorkVec(i,s)$, where $s$ is an input media unit.
\end{itemize}

To calculate agreement between 2 workers on the same media unit, we compute the cosine similarity over the 2 worker vectors. In order to reflect the dependency of the agreement on the degree of clarity of the annotations, we compute $Wcos$, the weighted version of the cosine similarity. The Annotation Quality Score (AQS), which will be described in more detail at the end of the section, is used as the weight. For open-ended tasks, where annotation quality cannot be calculated across multiple media units, we consider annotation quality equal to 1 (the maximum value) in all cases. Given 2 worker vectors, $vec_1$ and $vec_2$ on the same media unit, the formula for the weighted cosine score is:

\begin{align*}
& Wcos(vec_1, vec_2) = \\
& = \dfrac{\sum_{a} vec_1(a) \; vec_2(a) \; AQS(a)}{\sqrt{(\sum_{a} vec_1^2(a) \; AQS(a)) \; (\sum_{a} vec_2^2(a) \; AQS(a))}},
\end{align*}
$$\forall a \text{ - annotation}.$$

The {\bf Media Unit Quality Score (UQS)} expresses the overall worker agreement over one media unit. Given an input media unit $u$, $UQS(u)$ is computed as the average cosine similarity between all worker vectors, weighted by the worker quality ($WQS$) and annotation quality ($AQS$). Through the weighted average, workers and annotations with lower quality will have less of an impact on the final score. The formula used in its calculation is:

$$ UQS(u) = \dfrac{\sum\limits_{i, j} WorkVecWcos(i, j, u) \; WQS(i) \; WQS(j)}{\sum\limits_{i,j} WQS(i) \; WQS(j)}, $$
\begin{align*}
WorkVecWcos(i, j, u) = Wcos(&WorkVec(i, u), \\
                            &WorkVec(j, u)),
\end{align*}
$$\forall i, j \in workers(u), i \neq j.$$

The {\bf Worker Quality Score (WQS)} measures the overall agreement of one crowd worker with the other workers. Given a worker $i$, $WQS(i)$ is the product of 2 separate metrics - the worker-worker agreement $WWA(i)$ and the worker-media unit agreement $WUA(i)$:

$$ WQS(i) = WUA(i) \; WWA(i) .$$

The {\bf Worker-Worker Agreement (WWA)} for a given worker $i$ measures the average pairwise agreement between $i$ and all other workers, across all media units they annotated in common, indicating how close a worker performs compared to workers solving the same task.  The metric gives an indication as to whether there are consistently like-minded workers. This is useful for identifying communities of thought. $WWA(i)$ is the average cosine distance between the annotations of a worker $i$ and all other workers that have worked on the same media units as worker $i$, weighted by the worker and annotation qualities. Through the weighted average, workers and annotations with lower quality will have less of an impact on the final score of the given worker.

\begin{align*}
& WWA(i) = \\
& \dfrac{ \sum\limits_{j, u} WorkVecWcos(i,j,u) \; WQS(j) \; UQS(u) }{ \sum\limits_{j, u} WQS(j) \; UQS(u) }, \\
& \forall j \in workers(u \in units(i)), i \neq j .
\end{align*}

The {\bf Worker-Media Unit Agreement (WUA)} measures the similarity between the annotations of a worker and the aggregated annotations of the rest of the workers. In contrast to the $WWA$ which calculates agreement with individual workers, $WUA$ calculates the agreement with the consensus over all workers. $WUA(i)$ is the average cosine distance between the annotations of a worker $i$ and all annotations for the media units they have worked on, weighted by the media unit ($UQS$) and annotation quality ($AQS$). Through the weighted average, media units and annotations with lower quality will have less of an impact on the final score.

$$ WUA(i) = \dfrac{\sum\limits_{u \in units(i)} WorkUnitWcos(u, i) \; UQS(u)}{\sum\limits_{u \in units(i)} UQS(u)}, $$
\begin{align*}
 WorkUnitWcos(u, i) & = Wcos(WorkVec(i,u), \\
       &MediaUnitVec(u) - WorkVec(i, u))
\end{align*}

The {\bf Annotation Quality Score (AQS)} measures the agreement over an annotation in all media units that it appears. Therefore, it is only applicable to closed tasks, where the same annotation set is used for all input media units. It is based on $P_a(i | j)$, the probability that if a worker $j$ annotates $a$ in a media unit, worker $i$ will also annotate it.

\begin{align*}
& P_a(i | j) = \frac{ \sum\limits_{u} UQS(u) \; WorkVec(i, s)[a] \; WorkVec(j, s)[a] }{ \sum\limits_{u} UQS(u) \; WorkVec(j, u)(r) }, \\
& \forall u \in units(i) \cap units(j).
\end{align*}

Given an annotation $a$, $AQS(a)$ is the weighted average of $P_a(i | j)$ for all possible pairs of workers $i$ and $j$. Through the weighted average, input media units and workers with lower quality will have less of an impact on the final score of the annotation.

\begin{align*}
AQS(a) &= \dfrac{ \sum\limits_{i,j} WQS(i) \; WQS(j) \; P_a(i | j) }{ \sum\limits_{i,j} WQS(i) \; WQS(j) }, \\
& \forall i, j \text{ workers, }  i \neq j .
\end{align*}

The formulas for media unit, worker and annotation quality are all mutually dependent. To calculate them, we apply an iterative dynamic programming approach. First, we initialize each quality metric with the score for maximum quality (i.e. equal to 1). Then we repeatedly re-calculate the quality metrics until each of the values are stabilized. This is assessed by calculating the sum of variations between iterations for all quality values, and checking until it drops under a set threshold $t$.

The final metric we calculate is the {\bf Media Unit - Annotation Score (UAS)} -- the degree of clarity with which an annotation is expressed in a unit. Given an annotation $a$ and a media unit $u$, $UAS(u, a)$ is the ratio of the number of workers that picked annotation $u$ over all workers that annotated the unit, weighted by the worker quality.

$$ UAS(u, a) = \dfrac{ \sum\limits_{i \in workers(u)} WorkVec(i,u)(a) \; WQS(i) }{ \sum\limits_{i \in workers(u)} WQS(i) }. $$

\section{Conclusion}

In this paper, we present ongoing work into the CrowdTruth metrics, that capture and interpret inter-annotator disagreement in crowdsourcing. Typically crowdsourcing-based approaches to gather annotated data use inter-annotator agreement as a measure of quality. However, in many domains, there is ambiguity in the data, as well as a multitude of perspectives of the information examples. The CrowdTruth metrics model the inter-dependency between the three main components of a crowdsourcing system -- worker, input data, and annotation.

We have presented the definitions and formulas of several CrowdTruth metrics, including the three core metrics measuring the quality of workers, annotations, and input media units. The metrics are based on the idea of the triangle of disagreement, expressing how ambiguity in any of the corners disseminates and influences the other components of the triangle. Because of this, disagreement caused by low quality workers should not be interpreted as the data being ambiguous, but also that ambiguous input data should not be interpreted as due to the low quality of the workers. The metrics have already been applied successfully to use cases in topic relevance~\cite{inel2018studying}, semantic frame disambiguation~\cite{dumitrache2018capturing} and relation extraction from sentences~\cite{dumitrache2017false}.


\bibliographystyle{aaai}
\bibliography{biblio}

\end{document}